\documentclass[openacc]{rstransa}

\titlehead{Research Article}

\begin{document}

\title{Dynamic modeling of coronal abundances during flares on M-dwarf stars }

\author{
David H. Brooks$^{1,2}$, Jeffrey W. Reep$^{3}$, \\
Andy S.H. To$^{4}$, Luke Fushimi Benavitz$^{5}$ \\
and Lucas A. Tarr$^{6}$}

\address{$^{1}$Computational Physics, Inc., Springfield, VA 22151, USA\\
$^{2}$University College London, Mullard Space Science Laboratory, Holmbury St. Mary, Dorking, Surrey, RH5 6NT, UK\\
$^{3}$Institute for Astronomy, University of Hawai'i, Pukalani, HI 96768, USA\\
$^{4}$ESTEC, European Space Agency, Keplerlaan 1, PO Box 299, NL-2200 AG Noordwijk, Netherlands\\
$^{5}$Institute for Astronomy, University of Hawai'i, Honolulu, HI 96822, USA\\
$^{6}$National Solar Observatory, 22 Ohi'a Ku St., Makawao, HI 96768, USA}  

\subject{astrophysics, stars, spectroscopy}

\keywords{Solar corona, elemental abundances, solar flares, solar-stellar connection}

\corres{David H. Brooks\\
\email{dhbrooks.work@gmail.com}}

\begin{abstract}
Solar atmospheric elemental abundances are now known to vary both in space and time. Dynamic modeling
of these changes is therefore necessary to improve the accuracy of radiative hydrodynamic simulations. 
Recent studies have shown that including spatio-temporal variations in coronal abundances during
solar flares leads to the formation of coronal condensations (rain), which are otherwise
difficult to create in impulsively heated field aligned hydrodynamic flare models. These simulations start with a solar corona
dominated by the first ionization potential (FIP) effect, and evaporate photospheric material into the 
post-flare loops. 
We here explore perhaps the most extreme non-solar starting condition
for the coronal composition in these simulations: an initial corona dominated by the inverse FIP (iFIP) effect, such as 
is observed on active M-dwarf stars. We show that a flaring event in a corona enriched
with high FIP elements leads to a solution similar to the solar case. Coronal rain is harder to form 
by this method during
flares on M-dwarfs, however, if the corona is depleted of low FIP elements.
\end{abstract}

\begin{fmtext}
\end{fmtext}
\maketitle

\section{Introduction}
The solar corona is sustained by mass and energy flow from the lower atmospheric layers, so the elemental composition
was originally expected to be the same as that of the photosphere. Analysis of UV spectra, however,
showed that the abundances of Fe, Si, and Mg, are closer to the values seen in planetary nebulae and cosmic rays
\cite{Pottasch1963}. In fact, the slow solar wind also shows what is now known as the FIP (First Ionization Potential)
effect \cite{Meyer1985} i.e. a relative overabundance by a factor of 2-3 compared to the photosphere, of elements with a low FIP (less than $\sim$10 eV).
Recent observations from the {\it Hinode} EUV Imaging Spectrometer (EIS) \cite{Culhane2007}
have also highlighted that plasma composition can vary significantly both spatially and temporally
in many coronal features. See, for example, composition measurements over the full solar disk \cite{Brooks2015,Mihailescu2022}, in locations in post-flare loops
where coronal rain becomes noticeable \cite{Brooks2024}, or
unusual X-shaped flare ribbons \cite{To2021} and current sheets \cite{Warren2018}, and even the detection of the inverse FIP (iFIP) effect
- the depletion of low FIP elements or enhancement of those with a high FIP ($>$10 eV) -
in localized regions of sunspots during solar flares \cite{Doschek2015}.

Similar abundance variations are seen in spatially unresolved observations of solar-like stellar coronae. A previous study
reported a strong correlation between the observed FIP effect and the stellar spectral type \cite{Wood2010}. That is,
there is a progression from a solar-like FIP effect in G-type stars, through a smaller (or no) effect in K-type stars, to a strong
iFIP effect in M-dwarfs. There may also be variations with the activity cycle, as observed on the Sun-as-a-star \cite{Brooks2017}.
For further discussion, see also the review in this issue \cite{Brooks2026} for references
on changes over widely variant timescales from months (active regions) to years (solar cycle).

Large enhancements and spatio-temporal variations in element abundances make a significant difference to the radiative cooling properties
of the coronal plasma \cite{Cook1989}, and therefore can potentially impact numerical modelling and comparisons with observations.
This is because increases (or decreases) in elemental abundances, regardless of
ionisation potential, lead
to stronger (or weaker) radiation from spectral lines emitted by ions of those species, which in turn results in a greater
(or lesser) contribution to the total radiative losses. If the magnitude of the radiative losses is larger (or smaller)
this leads to faster (or slower) cooling.

There have been some recent advances made in studying these impacts.
To investigate the inverse FIP (iFIP) effect in solar flares, different radiative loss
functions reflecting the enhancement/depletion of high/low FIP elements were added as look-up tables to the (Enthalpy Based
Thermal Evolution of Loops) EBTEL 
\cite{Klimchuk2008,Cargill2012} 0-D
hydrodynamic code \cite{Brooks2018}, and analysis of the resultant post-flare loop cooling times showed that the low FIP elements
were likely depleted. A similar simulation setup was used to match simulated loop cooling times with the observed cooling times
of different locations in a post-flare loop arcade with known (measured) abundances \cite{Mihailescu2025}. In both these cases,
the starting (static) radiative loss function was altered to look for diagnostic differences in the cooling times, but the
studies were not based on a physical model for the abundance changes.

The EBTEL++ code \cite{Barnes2016,Barnes2025} has now been updated to incorporate time-variable abundances as a weighted average between 
the initial assumed coronal value and the material carried into the corona via chromospheric evaporation \cite{Reep2024}. More recently,
the 1-D HYDrodynamics and RADiation (HYDRAD) code \cite{Bradshaw2003,Bradshaw2013} has been modified to include an abundance factor
that is variable in both space and time \cite{Benavitz2025}. The study also looked at the formation of coronal rain. 

Coronal rain has been observed in UV/EUV data since at least the 1970s (see e.g. \cite{Ionson1978}, and references therein),
but was only studied occasionally \cite{Schrijver2001,DeGroof2005}. In recent years its importance has become more widely recognised
and it has become intensively studied observationally \cite{Antolin2010,Antolin2012,Kleint2014,Froment2015,Kohutova2016}.
Coronal rain is thought to consist of cool, dense plasma condensations that form due to a thermal instability when coronal loops cool catastrophically
\cite{Muller2003}. 
It appears in 3-D radiative MHD simulations when the thermal instability is triggered by chromospheric evaporation
in response to heating associated with magnetic field line braiding \cite{Kohutova2020}. It can also be produced in 1-D 
hydrodynamic simulations if the heating is localized to the loop footpoints. In fact there are several avenues
to producing these instabilities including the effects of flows and magnetic topology. See \cite{Keppens2025} for a recent
comprehensive review.

One outstanding issue is that it is difficult to produce coronal rain in electron beam heating
simulations of flare loops that assume a constant, or monotonically decreasing, magnetic field strength with height \cite{Reep2020}.
\cite{Benavitz2025} performed two loop simulations of typical short-duration
impulsive heating (nanoflares) and strong electron beam heating (flares) in active region loops. They found that
including spatio-temporally varying 
abundances causes coronal loop condensations (coronal rain) to form at the loop apex due to a localized peak in the radiative losses
relative to other values along the loop. 
This peak results from the localized enhancement in low FIP elements that forms because material evaporating from the chromosphere
pushes the coronal plasma to the apex.

Future work will determine the range of parameters necessary to produce coronal rain 
in typical solar flare conditions. An interesting question that arose during discussions of this work at the Theo Murphy meeting
was whether we could formulate
an extreme test of this simulation in quite different conditions from the solar case? In Sun-as-a-star time-series observations of solar flares,
the initial coronal abundance moves towards photospheric values due to chromospheric evaporation \cite{Warren2014,Mondal2021}. 
Similar behaviour is commonly seen in flares on solar analog stars, and in fact 
the basic picture of evaporation of photospheric composition plasma into the corona seems to hold regardless of the initial coronal composition since it
has also been seen on some stars with iFIP coronae in their quiescent state \cite{Nordon2008}. The latter would certainly represent a very different scenario 
from the solar case since the Sun and M-dwarfs sit at opposite extreme ends of the spectral type versus FIP effect correlation \cite{Wood2010}. 
To give some specific examples, increases from $\sim$0.3 and $\sim$0.6$\times$ solar photospheric 
to $\sim$1 and $\sim$1.4$\times$ photospheric were measured during flares observed on the eclipsing binary Algol using metals and the M6 variable
star CN Leo (Wolf 359) using Fe \cite{Favata1999,Liefke2010}. Increases in Fe and Si abundances from values below those of the solar photosphere 
were also observed during a flare on the RS CVn binary HR 1099 using XMM-Newton \cite{Audard2001}, and this has been seen again very recently,
for Fe and Ca during a flare, in HR 1099 observations from the Resolve instrument on XRISM \cite{Kurihara2025}.

Here we model the response of an iFIP corona in an M-dwarf star, to strong electron beam heating of the stellar chromosphere during a flare. Based on 
measurements of coronal abundances as a function of stellar spectral type, the simulations are broadly applicable to the coronae we expect
in M-dwarf stars \cite{Wood2010,Seli2022}. 
Since our goal here is to assess the impact of adopting an iFIP corona as the starting condition, we decided to control the experiment
by closely matching the rest of the model parameters from \cite{Benavitz2025}, and simulating a moderate solar flare. Flares of this magnitude 
should occur more frequently on M-dwarfs, but the distribution of flare energies spans a much larger range than on the Sun \cite{Kowalski2024}. Our simulation therefore
likely represents a flare from the cooler end of the M-dwarf distribution.
We simulate two iFIP scenarios: 1) the enrichment of high FIP ($>$10 eV) elements in the corona,
and 2) the depletion of low FIP elements in the corona. Although these two possibilities exist in active stars, evidence from theoretical
modeling \cite{Laming2004,Laming2015,Martinez2023}, observations of solar post-flare loop cooling times \cite{Brooks2018} and column depths 
\cite{Doschek2016}, and the detection of the iFIP effect in the slow solar wind \cite{Brooks2022}, all suggest that the FIP effect operates
to deplete low FIP elements, so this may be the more likely scenario. 

\section{Loop simulations}
We use HYDRAD
for our calculations, including optically thick chromospheric radiation \cite{Carlsson2012} and optically thin 
radiative losses computed using the CHIANTI database v.10 \cite{Dere1997,DelZanna2021}, 
assuming ionisation equilibrium within HYDRAD for the 15 most abundant elements. 
We also adopt the photospheric abundances of \cite{Asplund2009}.
HYDRAD is open source and is publically available
\footnote{https://github.com/rice-solar-physics/HYDRAD}.
The flare simulation setup follows the electron beam 
heated case of \cite{Benavitz2025} with two modifications. 
In general, coronal loop lengths are significantly larger 
on M-dwarfs \cite{Mullan2006}, so we adopt a loop length of 150 Mm. The other modification is a reduced heating duration.
We initiate the electron beam at the start of the simulation with a constant flux
of 2$\times$10$^{10}$ erg cm$^{-2}$ s$^{-1}$ for 10 s, a cutoff energy of 15 keV, and a spectral index of 5. 
Adaptive mesh refinement is employed to ensure adequate spatial resolution in the transition region,
with each grid cell allowed to split up to 12 times.

Our treatment of spatial and temporal abundance changes also uses the HYDRAD implementation of \cite{Benavitz2025}. This implementation
is also publically available 
\footnote{https://github.com/jwreep/HYDRAD/tree/Abundances} and modifications specific to this work have been placed in a Zenodo archive \cite{Reep2025}. 
The radiative loss rate within the simulation is modified by an abundance factor $f(t,s)$ that is variable in space, $s$, and time, $t$.
$f$ evolves due to flows along the loop following the advection equation 
\begin{equation}
 \frac{\partial f}{\partial t} + v \frac{\partial f}{\partial s} = 0.
\end{equation}
where $v$ is the bulk flow velocity. In what follows, we add a subscript $H$ or $L$ to be clear whether high or low FIP elements 
are being modified.

We perform two simulations of a flare occuring in an iFIP dominated stellar corona. As discussed in the introduction,
solar observations \cite{Doschek2016,Brooks2018,Brooks2022} and models \cite{Laming2004,Laming2015,Martinez2023} currently favor the idea that low FIP elements are depleted in the solar iFIP effect, but even if correct this
may not be the case in very different stellar coronae. We therefore explore two possible scenarios.
For the first simulation we 
assume the corona is initially enriched with 
high FIP elements by a factor of 4 and define $f_H = 4$ in the corona. 
In our second simulation we 
assume that the low FIP elements are depleted and 
set $f_L = 0.25$ in the corona.
In both cases we set $f = 1$ (photospheric abundances) in the chromosphere. Both initial conditions were implemented with a step 
function between the photopheric and coronal values at a distance of 2.26 Mm (the base of the transition region) from both ends of the loop.

\section{Results}
\begin{figure}[!h]
\centering\includegraphics[viewport=20 60 980 500,clip,width=\textwidth]{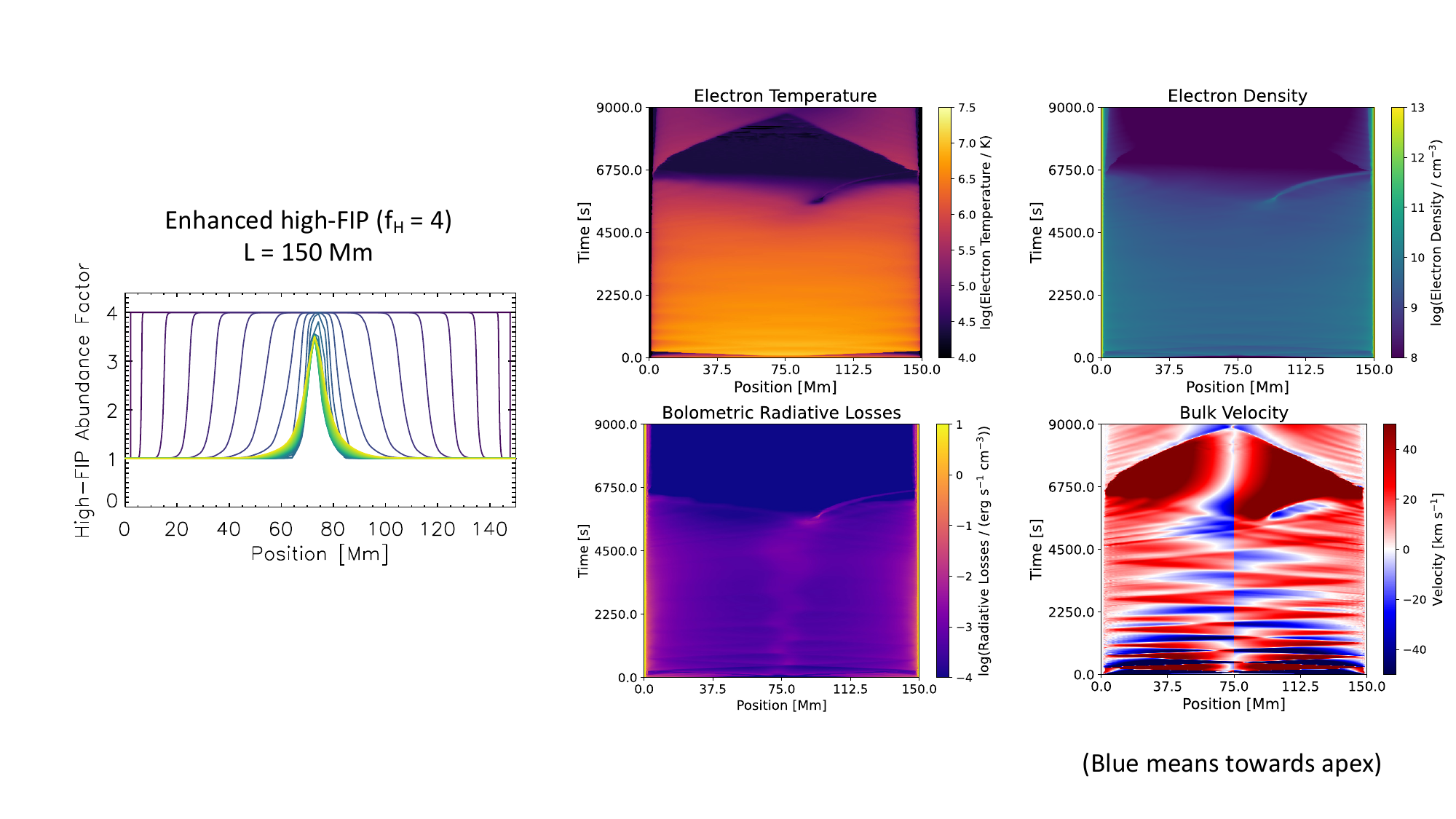}
\caption{ Electron beam heated loop simulation for the case where high FIP elements are enhanced in the initial corona.
Left panel: abundance factor as a function of position along the loop. Time is coded by color (purple$\rightarrow$green$\rightarrow$yellow) in 10 s
increments for the first 500 s of the simulation i.e. approximately the initial 17\% of the total simulation time shown in the subsequent panels.
Middle panel: electron temperature (top) and bolometric radiative losses (bottom) as a function of time along the loop.
Right panel: electron density (top) and bulk flow velocity (bottom) as a function of time along the loop.
Red/blue velocities indicate flows away from/towards the loop apex.  
The corona is initially enriched with high FIP elements by a factor of 4 (iFIP effect), 
but the abundances decrease rapidly towards photospheric values ($f_H$=1.0)
due to chromospheric evaporation as the loop heats up. This is similar to the simulation shown in \cite{Benavitz2025}. A localized peak in $f_H$ forms 
at the loop apex, which produces an associated increase in radiative losses and a faster cooling time.
A coronal rain condensation forms.
}
\label{fig1}
\end{figure}

The results of our simulation starting from a corona enriched with high FIP elements is shown in Figure \ref{fig1}.
Due to the broadly similar
temperature coverage of low and high FIP elements, the enhancement of high FIP elements
to a factor of $f_H = 4$ results in a radiative loss function that is expected to be broadly similar to the one that would be obtained
if the low FIP elements were enhanced instead. This is indirectly confirmed in Figure \ref{fig1} where the 
results are similar to the solar case presented by \cite{Benavitz2025}. 

\begin{figure}[!h]
\centering\includegraphics[viewport=20 60 980 500,clip,width=\textwidth]{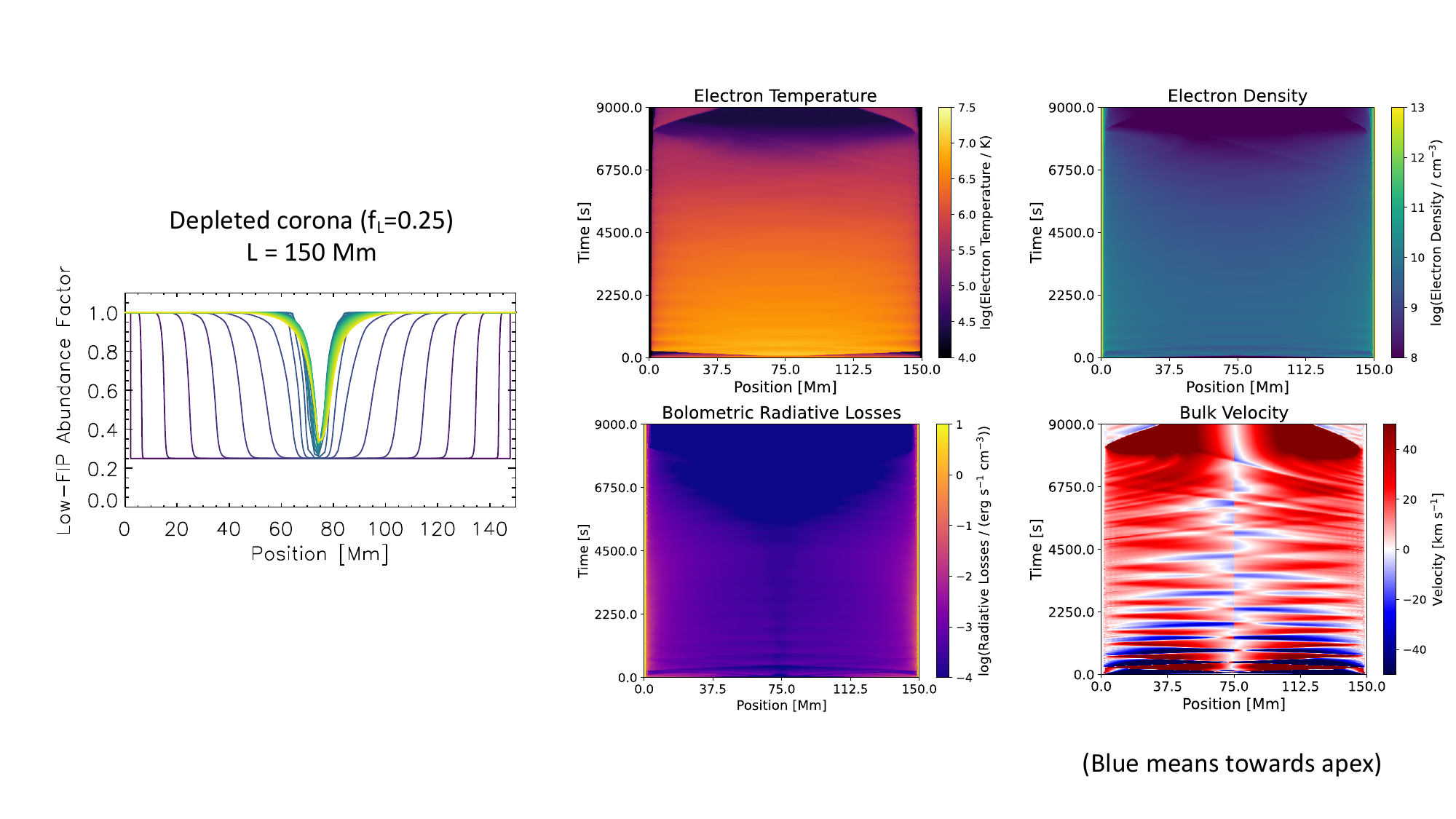}
\caption{ Same as Figure \ref{fig1} but for the case where low FIP elements are depleted in the initial corona.
The corona is initially depleted of low FIP elements by a factor of 4 (iFIP effect), but the abundances increase rapidly towards photospheric values ($f_L$=1.0)
due to chromospheric evaporation as the loop heats up. This is opposite to the simulation shown in Figure \ref{fig1} and in
\cite{Benavitz2025}. A localized dip in $f_L$ forms 
at the loop apex, which produces an associated decrease in radiative losses and a longer cooling time.
No coronal rain condensation forms.
}
\label{fig2}
\end{figure}

Broadly speaking, impulsive heating simulations produce a loop that heats up uniformly and then cools down \cite{Reep2020}.
The overall cooling time of the loop is dictated by the cooling timescales of radiation, conduction, and enthalpy fluxes 
\cite{Cargill1995,Bradshaw2005,Bradshaw2010,Cargill2013}. The radiative cooling timescale depends on the ratio of thermal
energy ($\propto n$) to the radiative loss rate ($\propto n^2$) and is therefore
affected by both the density ($\propto 1/n$) and the abundances that dictate the slope of the radiative loss curve, which 
dominates the cooling after the density peaks following evaporation.

In our simulation, once chromospheric evaporation in response to the electron
beam heating begins, material flows into the loop from the photospheric footpoints. 
The abundance factor then starts to decrease rapidly along the loop. This pushes plasma towards the loop top to
produce a localized peak in high FIP element abundances (factor $f_H$) and an associated spike in the radiative losses. Although the loop 
has initially heated up in a uniform manner, the localized peak in radiative losses causes enhanced cooling
at the loop apex leading to a local decrease in pressure and the formation of a pressure gradient that feeds the region with plasma 
and raises the density at the apex. This then exacerbates
the runaway radiative cooling and forms a coronal condensation. One consequence of the longer loop length in the simulation is that chromospheric 
evaporation takes longer to fill the loop and the condensation forms later than in \cite{Benavitz2025}. They found that the condensations formed
after $\sim$2000s, whereas here it takes more than 5000 s. The condensation is also not located exactly at the apex. The result, however, confirms that an iFIP corona enriched
with high FIP elements can lead to the formation of coronal rain during stellar flares.

The results of our simulation starting from a corona depleted of low FIP elements are shown in Figure \ref{fig2}.
In this case, the depletion factor of $f_L = 0.25$ produces a radiative loss function that has a smaller magnitude than the solar photospheric one.
See Figure 1 of \cite{Brooks2018} for an example. 
As can be seen in Figure \ref{fig2}, photospheric material flows into the loop 
due to chromospheric evaporation, and causes a rapid increase in the abundance factor along the loop.
As in the previous simulation, an accumulation of material with the initial coronal abundances forms at the loop top. This time, 
a localized dip in abundances of low FIP elements (factor $f_L$) is produced with an associated reduction in the radiative losses. 
The localized decrease in radiative losses causes reduced cooling
at the loop apex, so the loop effectively heats up and cools down in a uniform manner
without any formation of a condensation, as in simulations with fixed radiative losses \cite{Reep2020}. Coronal rain is therefore difficult
to form by this method in an iFIP corona depleted of low FIP elements, at least for these model conditions.

\section{Discussion}
Measurements of elemental abundances in the solar upper atmosphere have revealed several important issues. Plasma composition varies
spatially between different regions and features, and also varies as a function of time. Such variations 
modify theoretical radiative loss functions, and this in turn affects the cooling rate of the coronal plasma. 
Elemental abundance changes should therefore be incorporated into radiative hydrodynamic and magnetohydrodynamic models for future studies. 

Previous work has highlighted the impact of including spatio-temporally varying abundances in 
simulations of electron beam heating in flares \cite{Benavitz2025}.
It was shown that the evaporation of chromospheric plasma into post-flare loops leads to a localized peak in the radiative loss
rate that allows coronal condensations to form. This has been difficult to produce in previous field aligned hydrodynamic simulations of electron beam heating in flare loops that assume a constant magnetic field strength with height.

Here we have performed two simulations to further demonstrate the impact of elemental abundances that vary in space and time. We
set up a quite extreme contrasting example to the previous study that attempts to mimic the coronal composition found on active M-dwarf stars. 
These stars possess iFIP quiescent coronae, very different from the Sun, but as discussed in the introduction, stellar observations suggest that photospheric material is evaporated during stellar flares.

Observations suggest that the solar iFIP effect is due to the depletion of low FIP elements, but this may not be the case in 
the very different conditions on M-dwarf stars. An alternative possibility is that the iFIP effect there is due to the increase of high FIP elements. 
We therefore modeled both scenarios.
The enrichment of high FIP elements leads to a result that is similar to the solar case. The depletion of low FIP elements, however,
leads to reduced cooling at the loop apex, so the thermal runaway that causes condensations to form does not occur.

Our results suggest that coronal rain is less likely to form due to electron beam heating during flares on M-dwarfs. We should note that there are
significant differences on these stars compared to the solar case, however. 
Our simulations take into account one aspect of these differences: the expected longer loop lengths \cite{Mullan2006}. 
This apparently makes it even harder for chromospheric evaporation to form condensations.
Conversely, the heating rates would be expected to be higher and that could potentially cause faster evaporation. One simplification is
that our symmetric loop setup is somewhat idealised. In this issue, \cite{Reep2026} have shown that asymmetric heating affects whether and where
coronal rain can form. In particular, heating that is half as strong in one leg compared to the other leg does not result in rain formation.
This is because the localized peak in radiative losses at the apex is dissipated. 
We have verified this is also true in the longer 150 Mm loop case. If the heating rate is much stronger, however, or the elemental composition 
at the loop footpoints is also asymmetric, coronal rain can form and it may appear further down the loop leg.
This competition
between loop length, loop geometry, and heating rate will be further tested in future.
Furthermore, our radiation modelling assumes ionization equilibrium, an optically thin corona, and solar photospheric abundances.
Relaxing these assumptions is another avenue to explore. We have included all of our simulations in the Zenodo archive \cite{Reep2025} for interested readers (including those not shown here).

Notwithstanding these issues, and the exotic nature of some solar-like stars, we expect that the basic physics of the HYDRAD 
model is applicable to M-dwarfs. That said, it would be challenging to observationally resolve coronal rain condensations on these stars.
It is worthwhile, then, to reflect on what other aspects of these simulations might be detectable in stellar observations. One potential
diagnostic is the cooling rate. Despite the fact that both simulations evaporate photospheric material during the impulsive
event, the initial cooling function is significantly different. This implies that the cooling behaviour would also be different.
Considering the initial cooling functions alone, if high FIP elements are enhanced, we expect a transition from faster to slower cooling, whereas if low FIP 
elements are depleted, we expect a transition from slower to faster cooling. Of course the modeling is complex. The coronally averaged radiative losses
in our two simulations initially track each other quite closely. At later times, when
the condensation can form, we see diverging behaviour and multiple changes in which case cools faster or slower at any given time. 
This behaviour could be detectable in the light curves of
different EUV wavelength ranges, or X-ray energy bins, covering spectral lines formed at different temperatures. 
It has been argued that coronal rain may be seen in optical light curves from the Transiting Expolanet Survey Satellite (TESS) in the flare late-phase
\cite{Yang2023}.
Also, for
X-ray instruments observing energy ranges dominated by Fe emission, for example, we ought to only detect abundance variations
if the low FIP elements are depleted. As discussed, snapshots of abundance variations have been detected in high resolution spectra from
XRISM \cite{Kurihara2025}, and, despite the associated model fit uncertainties, also in time-resolved observations from NICER \cite{Kurihara2024}.

\vskip6pt
\ack{We thank Miki Kurihara for helpful discussions on stellar X-ray observations. We also thank the Royal Society for its support throughout the organization of the Theo Murphy meeting and the production of this special issue. The work of DHB was performed under contract to the Naval Research Laboratory and was funded by the NASA Hinode program. 
CHIANTI is a collaborative project involving George Mason University, the University of Michigan (USA), University of Cambridge (UK) and NASA Goddard Space Flight Center (USA).
}

\end{document}